%% file: template.tex
\documentclass[cameraready]{Interspeech}

\usepackage{cite}  
\usepackage{booktabs,tabularx,multirow,ragged2e,siunitx}
\usepackage{graphicx}
\usepackage{amsmath}
\usepackage{amssymb}  
\usepackage{tikz}
\usepackage{tabularx}
\usepackage{capt-of}
\usepackage{placeins}   
\usepackage{afterpage}       
\usepackage{afterpage}
\usepackage{dblfloatfix} 
\usepackage{xurl}
\usepackage{microtype}
\usepackage{threeparttable}
\usepackage{tablefootnote}
\usepackage{etoolbox}
\usepackage{booktabs}
\usepackage[table]{xcolor}

\newcommand{\sig}{\rlap{\textsuperscript{\textcolor{bmldAccent}{$\ast$}}}}

\title{Spectro-Temporal Interference Confounds Phase Encoding\\in Spatial Audio Foundation Models}

\author[affiliation={1,2}, orcid=0009-0009-0389-8307, equalcontribution, correspondingauthor]{Yuxuan}{Chen}
\author[affiliation={1,3}, orcid=0009-0007-9908-0504, equalcontribution]{Haoyuan}{Yu}
\author[affiliation={4}, orcid=0009-0008-3693-5576, equalcontribution]{Peize}{He}

\address{
    $^1$ The Chinese University of Hong Kong, Shenzhen, \\
    $^2$ Jilin University, 
    $^3$ Hunan University, \\
    $^4$ University of Electronic Science and Technology of China
}

\email{yxchen5522@mails.jlu.edu.cn, y15352176976@hnu.edu.cn, 2023300904027@std.uestc.edu.cn}


\begin{document}

\maketitle

\begin{abstract}
Recent spatial self supervised audio models achieve high performance on localization tasks, raising questions about their encoding of microsecond interaural phase fine structures. We propose a psychoacoustic benchmark based on the binaural masking level difference to evaluate this. Using an equalization cancellation baseline and a GCC PHAT positive control we evaluate nine frozen audio models spanning binaural SSL, monaural SSL, and neural audio codecs. Four monaural negative controls yield zero BMLD confirming binaural specificity. Two general purpose binaural SSL models exhibit minimal phase sensitivity while dedicated binaural spatial SSL models achieve substantial but sub-ceiling BMLD. Progressive physical ablations show that general purpose binaural SSL models rely on spectro temporal interference textures rather than cross channel phase computation. High detection rates in speech reflect a confounding reliance on broadband envelopes rather than genuine phase encoding.
\end{abstract}

\section{Introduction}
\label{sec:intro}

Spatial self-supervised audio models demonstrate high accuracy on macroscopic sound source localization tasks~\cite{wavjepa25,gram,soundspaces2,spatialhubert}. Existing benchmark evaluations predominantly assess geometric direction-of-arrival estimation and room acoustic properties~\cite{biseldhrtf}. These evaluation protocols do not verify whether models extract the microsecond interaural phase cues that govern spatial hearing~\cite{klumpp56itd}; even dedicated auditory motion benchmarks reveal systematic spatial deficits in current models~\cite{audiomotionbench}. Recent evidence shows that downstream task success does not guarantee human-like internal mechanisms~\cite{saddler24,tuckute23}, and that deep networks are prone to shortcut solutions that exploit superficial statistical regularities~\cite{geirhos20shortcut}. This leaves open the question of whether current spatial models encode genuine cross-channel phase information or rely on simpler heuristics.

The binaural masking level difference (BMLD) provides a standardized psychoacoustic test for cross-channel phase encoding~\cite{hirsh48,licklider48,bmldpuretone,vandepar99}. Human listeners improve detection thresholds by 10--17\,dB at low frequencies when the interaural phase of a target is inverted relative to a correlated noise masker~\cite{durlach63,colburndurlach78}. This spatial release relies on precise phase locking, which in humans is limited to frequencies below approximately 1.5\,kHz~\cite{brughera13itd}. Smith and Akeroyd applied supervised deep learning to predict BMLD psychometric functions~\cite{smith20bmld}. In contrast, we ask whether BMLD-like sensitivity emerges in frozen self-supervised binaural models without any task supervision.

We evaluate nine frozen models against an analytic equalization-cancellation (EC) baseline~\cite{durlach63} and a GCC-PHAT positive control. The evaluation spans binaural SSL models (WavJEPA~\cite{wavjepa25}, GRAM-T~\cite{gram}, Spatial-AST~\cite{spadavecchia24bat}, DSpAST~\cite{wilkinghoff2026dspast}), monaural SSL negative controls (HuBERT-Large~\cite{hsu21hubert}, WavLM-Large~\cite{chen22wavlm}, Wav2Vec2-Large~\cite{baevski20wav2vec2}), and neural audio codecs (EnCodec~\cite{defossez23encodec}, DAC~\cite{kumar24dac}). A bit-exact noise sharing protocol eliminates physical variance across conditions. Progressive physical ablations including high-pass filtering, energy nulling, and temporal fine structure vocoding isolate the detection mechanism. The data demonstrate that general-purpose binaural SSL models rely on per-channel spectro-temporal interference textures rather than cross-channel phase computation, while dedicated binaural spatial models achieve genuine phase sensitivity.

\input{tables/table1_model_comparison}

\section{Methodology}
\label{sec:method}

\subsection{Models and Feature Extraction}
\label{sec:models}

\noindent\textbf{Binaural SSL models.}
WavJEPA~\cite{wavjepa25} (12 transformer blocks, 768-dim, joint-embedding predictive architecture) and GRAM-T~\cite{gram} (12 transformer blocks, multi-channel masked autoencoder with spectrogram input) accept stereo waveforms. Spatial-AST is the binaural acoustic front-end of BAT~\cite{spadavecchia24bat} and uses explicit interaural phase difference (IPD) features. DSpAST~\cite{wilkinghoff2026dspast} is a later disentangled extension of Spatial-AST that introduces task-specific branches and feature attention over spatial audio features while building on the Spatial-AST/BAT framework.

\noindent\textbf{Neural audio codecs.}
EnCodec~\cite{defossez23encodec} is a binaural multi-codebook residual vector quantization codec that processes stereo input at 48\,kHz. DAC~\cite{kumar24dac} is a monaural audio codec serving as a negative control.

\noindent\textbf{Monaural SSL controls.}
HuBERT-Large~\cite{hsu21hubert}, WavLM-Large~\cite{chen22wavlm}, and Wav2Vec2-Large~\cite{baevski20wav2vec2} process only the left channel, providing strict negative controls since interaural phase manipulation is invisible to single-channel models.

All inputs are resampled to 16\,kHz and channel-normalized to unit variance. All models are evaluated frozen. The primary representation is the output of the final transformer block (or encoder output for codecs) followed by global average pooling.

\subsection{Stimuli and Control Protocol}
\label{sec:stimuli}

Synthetic targets are pure tones at center frequencies $f \in \{125, 250, 500, 1000, 2000, 4000\}$\,Hz. We construct three conditions per trial: diotic noise only ($N_0$), diotic tone in diotic noise ($S_0N_0$), and antiphasic tone in diotic noise ($S_\pi N_0$). Within each trial the masker waveform is bit-exact identical across all conditions, ensuring that any embedding difference is attributable solely to interaural phase manipulation. For ecological evaluation we render tones and speech excerpts from LibriSpeech~\cite{librispeech} through measured binaural room impulse responses from the AIR database~\cite{airdb}.
\subsection{Metrics and Baselines}
\label{sec:metrics}

We quantify representational spatial unmasking using a feature distance ratio converted to decibels:
\begin{equation}
\Delta_{\text{BMLD}} = 20\log_{10} \frac{\lVert Z(S_\pi N_0) - Z(N_0) \rVert_2}{\lVert Z(S_0 N_0) - Z(N_0) \rVert_2}
\label{eq:bmld}
\end{equation}
where $Z(\cdot)$ denotes the frozen model embedding. This representational metric complements perceptual binaural similarity measures such as BINAQUAL~\cite{binaqual} by directly probing internal model representations rather than rendered output quality.

\noindent\textbf{EC baseline.}
As a deterministic reference we implement the Durlach equalization-cancellation model~\cite{durlach63}. The cancellation residual yields a binaural signal-to-noise ratio
\begin{equation}
\text{SNR}_{\text{EC}} \propto \frac{P_S}{P_N \bigl[1 - \rho_{NLR}\, e^{-\frac{1}{2}(2\pi f \sigma_\tau)^2 - \sigma_\varepsilon^2}\bigr]}
\label{eq:ec}
\end{equation}
with masker interaural correlation $\rho_{NLR}$, internal time jitter $\sigma_\tau = 105\,\mu$s and dimensionless amplitude jitter $\sigma_\varepsilon = 0.25$~\cite{colburndurlach78}. The exponential decay in $f$ explains the characteristic low-frequency dominance of human BMLD.

\noindent\textbf{GCC-PHAT positive control.}
To verify physical phase integrity within the stimuli we compute the generalized cross-correlation with phase transform~\cite{knapp76gcc}:
\begin{equation}
\hat{R}_{12}(\tau) = \int \frac{G_{12}(f)}{|G_{12}(f)|}\, e^{\,j2\pi f\tau}\, df
\label{eq:gcc}
\end{equation}
The PHAT weighting $1/|G_{12}(f)|$ enforces spectral whitening so that the estimator depends exclusively on cross-channel phase alignment independent of amplitude or envelope structure. We compute GCC-PHAT on 4096-sample frames (256\,ms at 16\,kHz) with 75\% overlap.

\noindent\textbf{Statistical testing.}
Significance is assessed via sign-flip permutation tests ($\geq$5,000 iterations)~\cite{good05} with Benjamini--Hochberg false discovery rate correction ($q < 0.05$) applied across all cells within each model$\times$condition~\cite{benjamini95}. Bootstrap confidence intervals ($\geq$2,000 resamples)~\cite{efron93} are reported throughout. A \emph{significant cell} is defined as one frequency$\times$SNR combination whose 95\% bootstrap CI on the mean BMLD excludes zero after FDR correction. The number of seeds per cell varies across experiments and is stated in each table and figure caption.

\subsection{Physical Ablation Design}
\label{sec:ablation}

We implement a progressive set of physical ablations to isolate the driving detection mechanism. High-pass filtering above 2\,kHz removes all low-frequency content to test whether detection requires frequencies within the human phase-locking range~\cite{brughera13itd}. STFT and Mel domain energy equalization computes the short-time Fourier transform with a 2048-point Hann window and 512-sample hop. Within each time frame the root-mean-square energy is forced identical across the left and right channels independently in each of 128 Mel-spaced frequency bands, nulling macroscopic interaural level differences while preserving within-band fine structure. A 50\,Hz envelope vocoder~\cite{shannon95vocoder} extracts the Hilbert envelope below 50\,Hz per frequency band and re-modulates an uncorrelated Gaussian carrier, physically destroying the original temporal fine structure while preserving the macroscopic energy envelope.

\section{Results}
\label{sec:results}

\subsection{Absolute Representational Deficit}
\label{sec:deficit}

Table~\ref{tab:model_comparison} compares the representational BMLD at 500\,Hz across all nine neural models. At SNR\,=\,$-$14\,dB, the EC model provides +15.7\,dB of analytical separation. WavJEPA yields +0.5\,dB and GRAM-T yields +2.1\,dB, deficits of $\times$31 and $\times$7.5 relative to EC. The four monaural controls (HuBERT-Large, WavLM-Large, Wav2Vec2-Large, DAC) produce exactly 0.0\,dB BMLD across all conditions, confirming that the effect requires binaural input and is not a generic SSL artifact.

Spatial-AST reaches +6.8\,dB at SNR\,=\,$-$14\,dB, under half the +15.7\,dB EC reference rather than matching it. This demonstrates that a binaural SSL architecture with IPD-aware input features captures substantial, though incomplete, cross-channel phase information. DSpAST yields a comparable +7.0\,dB under the same conditions. EnCodec produces +7.0\,dB despite its codec objective function, suggesting that its binaural residual vector quantization preserves substantial interaural structure, matching the binaural SSL models.

\begin{figure*}[t]
  \centering
  \includegraphics[width=0.98\textwidth]{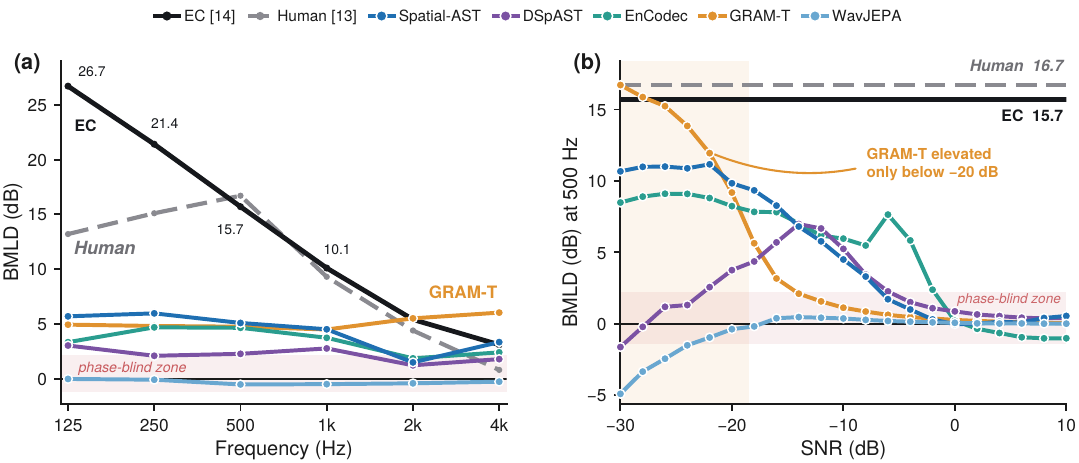}
  
  \caption{Absolute representational deficit. \textbf{(a)}~BMLD across
frequency for the EC analytical ceiling and the human threshold versus frozen neural models, all shown as full curves. \textbf{(b)}~BMLD at 500\,Hz versus SNR; the EC and
human references are SNR-invariant, and the neural curves are consistent with table~\ref{tab:model_comparison} at the tabulated SNRs.
Shaded band: phase-blind zone. Monaural controls are exactly 0\,dB.}

  \label{fig:deficit}
\end{figure*}

The frequency profiles diverge from EC predictions (Figure~\ref{fig:deficit}a). The EC baseline attenuates with increasing frequency, consistent with the exponential decay in Eq.\,\ref{eq:ec}. GRAM-T exhibits a frequency-invariant profile that does not follow this monotonic attenuation. WavJEPA remains below +0.5\,dB across all frequencies. The SNR dimension (Figure~\ref{fig:deficit}b) reveals that GRAM-T's elevated BMLD arises only at very low SNR ($< -20\,\mathrm{dB}$), collapsing toward zero at moderate SNR, whereas the EC baseline and human threshold are SNR-invariant.

\subsection{Temporal Pooling Artifacts}
\label{sec:pooling}

Temporal aggregation inflates weak per-frame effects. For WavJEPA, removing global average pooling causes the feature distance ratios to collapse into the phase-blind zone (mean ratio $< 1.05$); frame-wise computations show no significant spatial release. GRAM-T frame-wise ratios decrease but remain above unity. Pooled ratios exceed frame-wise ratios by a factor of $\times$2.1 for GRAM-T and $\times$3.4 for WavJEPA (Supplementary Figure~S1). These findings indicate that temporal averaging amplifies the weak per-frame effects reported in \S\ref{sec:deficit}.

\subsection{Macroscopic Cue Sensitivity}
\label{sec:cue}

\begin{figure}[t]
  \centering
  \includegraphics[width=0.96\linewidth]{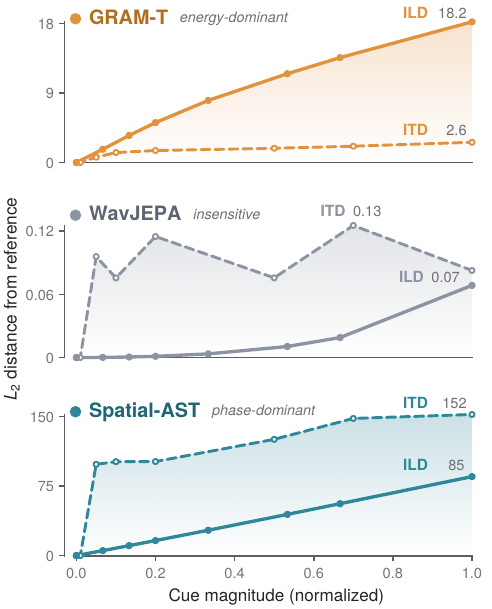}
  \vspace{-1mm}
  
\caption{\textbf{Cue sensitivity probes.} $L_2$ embedding distance from
the zero-cue reference vs.\ normalised cue magnitude (ILD $0$--$15$\,dB,
solid; ITD $0$--$1000$\,$\mu$s, dashed; $n{=}500$/pt). GRAM-T is energy-dominant ($18.2$ vs.\ $2.6$, $\times$7), WavJEPA insensitive to both with a noisy ITD response, and
Spatial-AST the inverse ($152$ vs.\ $85$, $\times$0.6),
confirming genuine phase encoding.}

  \label{fig:shortcut}
  \vspace{-4mm}
\end{figure}

Univariate acoustic cue probes isolate the features driving the neural representations (Figure~\ref{fig:shortcut}). GRAM-T feature distances scale monotonically with ILD over a 0--15\,dB range, reaching a peak of 18.2. The ITD probe yields a peak of approximately 2.6, a $\times$7 gap relative to the ILD response. WavJEPA is ILD-insensitive (peak $\approx 0$) and shows noisy, non-systematic ITD responses.

Spatial-AST exhibits the inverse pattern: ITD sensitivity (peak~152) exceeds ILD sensitivity (peak~85), yielding an ILD-to-ITD ratio of 0.6$\times$. This reversal relative to GRAM-T's 7$\times$ ratio confirms that IPD-aware binaural training produces a genuinely phase-dominant representation. The four monaural controls yield zero response to both ILD and ITD probes, as expected.

\subsection{Isolation of Interference Mechanisms}
\label{sec:isolation}

\begin{figure}[t]
  \centering
  \includegraphics[width=0.95\linewidth]{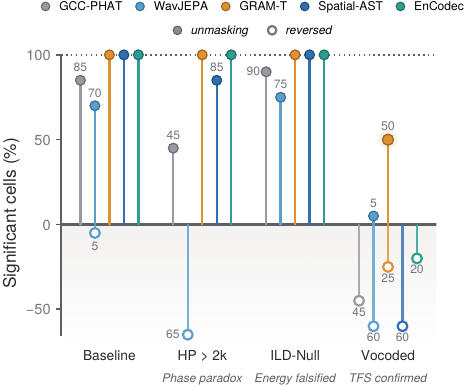}
  \captionsetup{font=small}
  
\caption{\textbf{Progressive falsification via physical ablations.}
Proportion of $n{=}20$ frequency\,$\times$\,SNR cells with significant
unmasking or reversal. GRAM-T and EnCodec resist high-pass filtering and Mel-band ILD nulling (100\%); only sub-50\,Hz vocoding collapses detection, and GRAM-T alone keeps a positive unmasking component (50\% of cells; 25\% reversed, 75\% significant overall)—an envelope-texture mechanism.}

  \label{fig:ablation}
  \vspace{-3mm}
\end{figure}

The progressive ablation results are summarized in Figure~\ref{fig:ablation}. The GCC-PHAT positive control maintains high detection rates under energy equalization (90\%), confirming the physical integrity of phase information within the stimuli.

High-pass filtering above 2\,kHz does not degrade GRAM-T detection rates, which remain at 100\%. Spatial-AST decreases from 100\% to 85\% under the same filter, consistent with the frequency dependence of biological ITD encoding~\cite{brughera13itd}. Human ITD sensitivity ceases above approximately 1.5\,kHz; GRAM-T's sustained performance beyond this limit indicates a mechanism independent of phase locking.

Under Mel energy equalization, GRAM-T maintains 100\% significance, EnCodec maintains 100\%, and Spatial-AST retains 100\%. This indicates that the remaining detection cue is distinct from macroscopic band energy across all model types.

Vocoder processing~\cite{shannon95vocoder} replaces the temporal fine structure with a decorrelated noise carrier. This is the only condition that differentially impairs the models: GRAM-T drops from 100\% to 75\%, EnCodec drops from 100\% to 20\%, and Spatial-AST drops from 100\% to 60\%. A systematic cutoff sweep (Supplementary Figure~S2) reveals a sharp recovery threshold at 32\,Hz matching the mel-spectrogram temporal resolution ($\sim$31\,ms frame rate), confirming that GRAM-T's operative cue resides in fast envelope textures rather than phase fine structure.

\subsection{Ecological Confound}
\label{sec:ecological}

\input{tables/table2_ecological}

Table~\ref{tab:ecological} reveals a consistent hierarchy under realistic speech conditions rendered through measured BRIRs. Spatial-AST and GRAM-T achieve 100\% significance in speech conditions, while EnCodec reaches 100\% and DSpAST 75\%. WavJEPA reaches 71\%. All four monaural controls yield 0\% significance, confirming that ecological detection requires binaural access.

The tone-to-speech improvement varies by mechanism: Spatial-AST improves from 83\% to 100\%, reflecting broadband activation of its IPD pathway. GRAM-T maintains 100\% via interference textures in both conditions. The stark 0\% monaural baseline excludes the possibility that BMLD significance arises from generic learned features unrelated to binaural input.

\section{Discussion}
\label{sec:discussion}

Pretraining loss functions and architectural constraints explain the observed representational properties. The masked autoencoder objective underlying GRAM-T~\cite{he22mae} optimizes L2 reconstruction of real-valued spectrogram patches, a loss dominated by high-energy envelope gradients that treats low-amplitude phase ripples as negligible residuals. The joint-embedding predictive architecture underlying WavJEPA~\cite{assran23ijepa} aligns latent representations across masked and target blocks, deliberately discarding pixel-level waveform detail to achieve semantic invariance. Neither loss function contains complex-domain terms or explicit phase alignment constraints, preventing the encoding of microsecond cross-channel interactions.

The expanded evaluation reinforces and contextualizes this conclusion. Four monaural negative controls produce zero BMLD uniformly, establishing that binaural access is necessary but not sufficient for phase encoding. EnCodec preserves some binaural structure through its residual vector quantization but its codec loss function provides no incentive for explicit phase alignment, yielding moderate BMLD that collapses under vocoding (100\%$\to$20\%). 
Conversely, Spatial-AST's IPD-aware binaural architecture achieves +6.8 dB at 500 Hz—well below the EC ceiling—demonstrating that genuine, if sub-ceiling, phase sensitivity is architecturally achievable within a binaural SSL framework.
Its vulnerability to high-pass filtering (100\%$\to$85\%) mirrors the biological frequency dependence of ITD encoding, further validating the benchmark. These findings are consistent with the hemispheric two-channel cancellation framework of Encke and Dietz~\cite{hemi2ch}, which shows that binaural unmasking in humans is well explained by a two-channel EC process operating below 1.5\,kHz.

The immunity to energy equalization combined with vulnerability to TFS vocoding reveals the operative mechanism. Phase inversion generates localized beating and micro-envelope ripples within short spectro-temporal units. Energy equalization nulls the macroscopic band energy but cannot eliminate the two-dimensional textures formed by these ripples across adjacent time-frequency patches. This pattern aligns with the sound texture perception framework of McDermott and Simoncelli~\cite{mcdermott11texture}. The ecological confound in realistic speech conditions arises because broadband signals interacting with head asymmetry naturally produce massive cross-channel envelope decorrelation, highly activating these texture-based heuristics. Binaural grouping models that rely on precisely these envelope and fine-structure cues for predicting spatial release from masking~\cite{binauralgrouping} further illustrate why coarse binaural access without genuine phase computation inflates ecological detection rates.

\section{Conclusion}
\label{sec:conclusion}

The proposed psychoacoustic benchmark, evaluated across nine frozen models spanning binaural SSL, monaural SSL, and neural audio codecs, reveals that general-purpose binaural SSL models rely on per-channel spectro-temporal interference features rather than cross-channel phase computation. Four monaural negative controls confirm binaural specificity, while Spatial-AST establishes that substantial, though sub-ceiling, phase sensitivity is achievable within a binaural SSL framework. Progressive physical ablations confirm the operative mechanism: high-pass filtering and energy nulling leave detection intact, while TFS vocoding degrades it. High detection rates in ecological downstream tasks reflect a confounding reliance on broadband envelope cues. These findings align with recent evidence that deep networks construct invariant feature manifolds fundamentally divergent from biological perception~\cite{feather23metamers}. Future spatial audio pretraining must incorporate explicit phase-aware constraints to bridge the gap between statistical pattern matching and genuine binaural computation.


\clearpage   
\nocite{*}   

\begingroup  
\raggedbottom

\section{Generative AI Use Disclosure}
We used a generative AI tool to assist with language editing and polishing of the manuscript, including improving grammar, clarity, and readability.
The tool was not used to generate scientific content, experimental results, or conclusions.
All coauthors reviewed the final manuscript and take full responsibility for it.

\bibliographystyle{IEEEtran}
\bibliography{mybib}
\endgroup   

\end{document}

%% file: tables/table1_model_comparison.tex
\begin{table}[t]
  \centering
  \definecolor{heatblue}{RGB}{33,102,172}                 
  \definecolor{ctrlgray}{gray}{0.93}                      
  \providecommand{\hb}{}\renewcommand{\hb}[2]{\cellcolor{heatblue!#1}#2}              
  \providecommand{\hh}{}\renewcommand{\hh}[1]{\cellcolor{heatblue!13}\textbf{#1}{\,\footnotesize dB}} 
  \providecommand{\sig}{}\renewcommand{\sig}[1]{#1\rlap{\textsuperscript{*}}}         
  \providecommand{\nsig}{}\renewcommand{\nsig}[1]{#1}                                 
  \providecommand{\zz}{}\renewcommand{\zz}{\nsig{\textcolor{black!55}{0.0}}}          
  \providecommand{\mbin}{}\renewcommand{\mbin}{\textcolor{heatblue}{$\bullet$}}       
  \providecommand{\mmono}{}\renewcommand{\mmono}{\textcolor{black!45}{$\circ$}}       
  \providecommand{\mk}{}\renewcommand{\mk}[1]{\makebox[0.95em][l]{#1}}                
  \providecommand{\gcite}{}\renewcommand{\gcite}[1]{{\footnotesize\textcolor{black!50}{\cite{#1}}}} 
  \providecommand{\sw}{}\renewcommand{\sw}[1]{\textcolor{heatblue!#1}{\rule[-0.3ex]{0.72em}{1.55ex}}}
  \providecommand{\heatscale}{}\renewcommand{\heatscale}{\mbox{0\,\sw{8}\sw{21}\sw{35}\sw{49}\sw{62}\,{+}13}}
  \caption[Cross-model BMLD comparison at 500\,Hz.]{Cross-model BMLD
  comparison at 500\,Hz. Columns: input SNR; cell values: BMLD gain in dB
  (Eq.~\ref{eq:bmld}). \mbin~binaural\enspace\mmono~monaural. Cell shading \heatscale\,dB $=$ gain. Sign-flip permutation tests
  with FDR correction.}
  \label{tab:model_comparison}
  \small
  \setlength{\tabcolsep}{2.5pt}
  \renewcommand{\arraystretch}{1.2}
  \begin{tabularx}{\linewidth}{ll*{3}{>{\centering\arraybackslash}X}}
    \toprule
    \rowcolor{gray!15}
    \textbf{Model} & \textbf{Type} & \hh{$-$14} & \hh{$-$4} & \hh{$0$} \\
    \midrule
    \mk{\mbin}EC~\gcite{durlach63} & Baseline
      & \hb{61}{\nsig{+15.7}} & \hb{61}{\nsig{+15.7}} & \hb{61}{\nsig{+15.7}} \\
    \midrule[0.03em]
    \mk{\mbin}Spatial-AST~\gcite{spadavecchia24bat} & SSL
      & \hb{32}{\sig{+6.8}} & \hb{5}{\sig{+1.0}} & \hb{0}{\nsig{$-$0.0}} \\
    \mk{\mbin}DSpAST~\gcite{wilkinghoff2026dspast} & SSL
      & \hb{33}{\sig{+7.0}} & \hb{7}{\sig{+1.5}} & \hb{4}{\sig{\textbf{+0.9}}} \\
    \mk{\mbin}EnCodec~\gcite{defossez23encodec} & Codec
      & \hb{33}{\sig{\textbf{+7.0}}} & \hb{28}{\sig{\textbf{+5.8}}} & \hb{1}{\sig{+0.2}} \\
    \mk{\mbin}GRAM-T~\gcite{gram} & SSL
      & \hb{10}{\sig{+2.1}} & \hb{2}{\sig{+0.5}} & \hb{1}{\sig{+0.3}} \\
    \mk{\mbin}WavJEPA~\gcite{wavjepa25} & SSL
      & \hb{2}{\sig{+0.5}} & \hb{1}{\sig{+0.1}} & \hb{0}{\nsig{+0.1}} \\
    \addlinespace[2pt]
    \rowcolor{ctrlgray}\mk{\mmono}Monaural\,$\times4^{\,\ddagger}$ & ---
      & \zz & \zz & \zz \\
    \bottomrule
  \end{tabularx}
  \par\vspace{1.2mm}
  \parbox{\linewidth}{\footnotesize\raggedright
  \textsuperscript{*}\,Significant after FDR correction ($q{=}0.05$);
  100 seeds/cell, 40 for monaural controls.
  \textsuperscript{$\ddagger$}\,Monaural controls (left channel only):
  DAC~\gcite{kumar24dac}, HuBERT-L~\gcite{hsu21hubert},
  WavLM-L~\gcite{chen22wavlm}, Wav2Vec2-L~\gcite{baevski20wav2vec2};
  all identically $0.0$\,dB ($0$ by construction).\par}
\end{table}

%% file: tables/table2_ecological.tex
\begin{table}[t]
  \centering
  \definecolor{heatblue}{RGB}{33,102,172}
  \definecolor{barblue}{RGB}{33,102,172}
  \definecolor{ctrlgray}{gray}{0.93}
  \newlength{\barw}\setlength{\barw}{0.9cm}
  \newlength{\barhh}\setlength{\barhh}{1.25ex}
  \newlength{\numw}\setlength{\numw}{1.95em}
  \newlength{\pbfill}\newlength{\pbrest}
  \providecommand{\barbox}{}\renewcommand{\barbox}[1]{%
    \setlength{\pbfill}{#1\barw}%
    \setlength{\pbrest}{\dimexpr\barw-\pbfill\relax}%
    \textcolor{barblue}{\rule[-.3ex]{\pbfill}{\barhh}}%
    \textcolor{black!8}{\rule[-.3ex]{\pbrest}{\barhh}}}
  \providecommand{\cellbar}{}\renewcommand{\cellbar}[2]{%
    \mbox{\barbox{#1}\hspace{0.4em}\makebox[\numw][r]{\footnotesize #2}}}
  \providecommand{\zc}{}\renewcommand{\zc}{%
    \mbox{\textcolor{black!8}{\rule[-.3ex]{\barw}{\barhh}}%
    \hspace{0.4em}\makebox[\numw][r]{\footnotesize\textcolor{black!45}{0/24}}}}
  \providecommand{\mbin}{}\renewcommand{\mbin}{\textcolor{heatblue}{\textbullet}}
  \providecommand{\mmono}{}\renewcommand{\mmono}{\textcolor{black!45}{\small$\circ$}}
  \providecommand{\mk}{}\renewcommand{\mk}[1]{\makebox[0.95em][l]{#1}}
  \providecommand{\gcite}{}\renewcommand{\gcite}[1]{{\footnotesize\textcolor{black!50}{\cite{#1}}}}
  \providecommand{\hh}{}\renewcommand{\hh}[1]{\cellcolor{heatblue!13}\textbf{#1}}
  \caption[Ecological confound under realistic stimuli.]{Ecological
  confound under realistic stimuli. Significant cells / 24
  frequency\,$\times$\,SNR conditions (95\% bootstrap CI excludes zero,
  $N{=}100$ seeds) under measured BRIRs. \mbin~binaural,
  \mbox{\mmono~monaural}; bar length~$=$~proportion significant.}
  \label{tab:ecological}
  \small
  \setlength{\tabcolsep}{2.5pt}
  \renewcommand{\arraystretch}{1.25}
  \begin{tabularx}{\linewidth}{ll*{2}{>{\centering\arraybackslash}X}}
    \toprule
    \rowcolor{gray!15}
    \textbf{Model} & \textbf{Type} & \hh{Tone} & \hh{Speech-like} \\
    \midrule
    \mk{\mbin}Spatial-AST~\gcite{spadavecchia24bat} & SSL
      & \cellbar{0.833}{20/24} & \cellbar{1.0}{24/24} \\
    \mk{\mbin}DSpAST~\gcite{wilkinghoff2026dspast} & SSL
      & \cellbar{0.625}{15/24} & \cellbar{0.75}{18/24} \\
    \mk{\mbin}EnCodec~\gcite{defossez23encodec} & Codec
      & \cellbar{0.75}{18/24} & \cellbar{1.0}{24/24} \\
    \mk{\mbin}GRAM-T~\gcite{gram} & SSL
      & \cellbar{1.0}{24/24} & \cellbar{1.0}{24/24} \\
    \mk{\mbin}WavJEPA~\gcite{wavjepa25} & SSL
      & \cellbar{0.583}{14/24} & \cellbar{0.708}{17/24} \\
    \addlinespace[2pt]
    \rowcolor{ctrlgray}\mk{\mmono}Monaural\,$\times4^{\,\ddagger}$ & ---
      & \zc & \zc \\
    \bottomrule
  \end{tabularx}
  \par\vspace{0.8mm}
  \parbox{\linewidth}{\footnotesize\raggedright
  \textsuperscript{$\ddagger$}\,DAC~\gcite{kumar24dac},
  HuBERT-L~\gcite{hsu21hubert}, WavLM-L~\gcite{chen22wavlm},
  Wav2Vec2-L~\gcite{baevski20wav2vec2}.\par}
\end{table}